\documentclass[aps,reprint,superscriptaddress]{revtex4-2}
\usepackage{amsmath}
\usepackage{amssymb}
\usepackage{graphicx}
\usepackage{orcidlink}
\usepackage{hyperref}
\usepackage{bm}
\usepackage{color}
\usepackage{pdfpages}

\hypersetup{colorlinks=true}

\makeatletter
\AtBeginDocument{\let\LS@rot\@undefined}
\makeatother

\newcommand{\setParDis}{\setlength {\parskip} {0.35cm}}
\newcommand{\setParDef}{\setlength {\parskip} {0.3pt}}

\hypersetup{
	colorlinks=true,
	citecolor=blue,
	urlcolor=magenta
}
\begin{document}
\title{Generation of Isolated Collimated Polarized $\gamma$-ray Beams via Spatiotemporal Optical Vortex Modulation}

\affiliation{State Key Laboratory of Ultra-Intense Laser Science and Technology, Shanghai Institute of Optics and Fine Mechanics (SIOM), Chinese Academy of Sciences (CAS), Shanghai 201800, China}
\affiliation{Laboratory of Zhongyuan Light, School of Physics, Zhengzhou University, Zhengzhou 450001, China}
\affiliation{State Key Laboratory of Dark Matter Physics, Key Laboratory for Laser Plasmas (MoE), School of Physics and Astronomy, Shanghai Jiao Tong University, Shanghai 200240, China}
\affiliation{Center of Materials Science and Optoelectronics Engineering, University of Chinese Academy of Sciences, Beijing 100049, China}

\author{Xinyu Xie}
\thanks{These authors have contributed equally to this work.}
\affiliation{State Key Laboratory of Ultra-Intense Laser Science and Technology, Shanghai Institute of Optics and Fine Mechanics (SIOM), Chinese Academy of Sciences (CAS), Shanghai 201800, China}
\affiliation{Center of Materials Science and Optoelectronics Engineering, University of Chinese Academy of Sciences, Beijing 100049, China}

\author{Fengyu Sun}
\thanks{These authors have contributed equally to this work.}
\affiliation{State Key Laboratory of Ultra-Intense Laser Science and Technology, Shanghai Institute of Optics and Fine Mechanics (SIOM), Chinese Academy of Sciences (CAS), Shanghai 201800, China}

\author{Huai-Hang Song}\email{huaihangsong@zzu.edu.cn}
\affiliation{Laboratory of Zhongyuan Light, School of Physics, Zhengzhou University, Zhengzhou 450001, China}

\author{Wei-Min Wang} \email{weiminwang1@sjtu.edu.cn}
\affiliation{State Key Laboratory of Dark Matter Physics, Key Laboratory for Laser Plasmas (MoE), School of Physics and Astronomy, Shanghai Jiao Tong University, Shanghai 200240, China}

\author{Wenpeng Wang}
\email{wangwenpeng@siom.ac.cn}
\affiliation{State Key Laboratory of Ultra-Intense Laser Science and Technology, Shanghai Institute of Optics and Fine Mechanics (SIOM), Chinese Academy of Sciences (CAS), Shanghai 201800, China}
\affiliation{Center of Materials Science and Optoelectronics Engineering, University of Chinese Academy of Sciences, Beijing 100049, China}

\date{\today}

\begin{abstract}
Attosecond, collimated, bright, polarized $\gamma$-ray sources are in high demand across nuclear physics, astrophysics, and high-energy physics. However, realizing attosecond duration, high collimation, high brilliance, and high polarization simultaneously within a single isolated source remains an outstanding challenge, owing to the inherent trade-offs between beam trapping and radiative dynamics. Here, we propose a novel scheme to generate an isolated, collimated, high-brilliance, polarized attosecond $\gamma$-ray beam from conventional solid foils irradiated by a linearly polarized spatiotemporal optical vortex (STOV) laser pulse accessible in Lab. Three-dimensional spin-resolved particle-in-cell simulations reveal that this relativistic-intensity STOV pulse can trap and accelerate electrons at its spatiotemporal singularity, producing a compact isolated electron bunch. This electron bunch subsequently undergoes head-on collision with the reflected laser pulse, which generates isolated $\gamma$-ray beams through nonlinear Compton scattering. With a peak intensity of $7\times10^{21}$ W/cm$^2$, we observe an isolated collimated ($\sim1.5^{\circ}$) $\gamma$-ray beam with an average linear polarization of $>60\%$ and a duration of $\sim$500 attoseconds. This approach is feasible with current or upcoming laser facilities and robust against variations in laser and target parameters, highlighting the capability of spatiotemporal structured light field modulation to address outstanding problems in plasma physics.
\end{abstract}

\maketitle

\setParDef
Polarization, as an intrinsic photon degree of freedom, endows polarized $\gamma$ photons with unique capabilities in nuclear physics~\cite{zilges2022,fagg1959RMP,Speth1981rpp}, high-energy physics~\cite{bragin2017prl,song2024pre,song2025njp,xue2023prl}, and astrophysics~\cite{woosley1993apj,Dean2008science,Gong2023PRL,Gong2025PRL}. These applications not only exploit polarization properties but also impose stringent demands on pulse duration, flux, brightness and collimation~\cite{Reichwein2025,Sun2022RMPP,Wang2025RMPP}. In nuclear physics, ultrafast processes such as nuclear single-particle transitions and resonance fluorescence on femtosecond timescales~\cite{zilges2022}, and resonant internal conversion on attosecond timescales~\cite{li2015prl}; these processes typically feature small cross sections (a few to tens of millibarns~\cite{Kossov2002}), and polarized $\gamma$ photons promise to significantly enhance the excitation efficiency~\cite{Speth1981rpp,Akbar2017PRC}. Realizing such studies thus demands $\gamma$ photons that simultaneously combine a high degree of polarization, MeV-level energies, and ultrashort pulse durations. Conventional radiation mechanisms, however, face fundamental bottlenecks in meeting all of these criteria: ultrafast radiation based on high-harmonic generation in gases provides excellent ultrashort pulse characteristics, yet photon energies remain confined to the extreme ultraviolet to soft X-ray regime~\cite{Sansone2011np,Yeung2017np,Teubner2009RMP}; polarized sources based on bremsstrahlung or linear Compton scattering can reach MeV energies but are limited in pulse duration and peak brilliance~\cite{Abbott2016prl,Ugg2005rmp,Petrillo2015prab,MAMI1994}.

Recently, the rapid advancement of high-power laser technology has enabled peak intensities to $10^{22}$--$10^{23}$ W/cm$^2$~\cite{Danson2019hpl,Zhang2020hpl,Yoon2021optica}, opening all-optical pathways to $\gamma$-ray generation via laser-plasma interactions and extending radiation energies to the MeV range with drastically enhanced peak brilliance~\cite{yan2017np,zhu2020sa,wang2018pnas,TaPhuoc2012np}. Nevertheless, simultaneously controlling multiple parameters, including attosecond pulse durations and polarization resolution remains challenging. Existing laser-plasma $\gamma$-ray source schemes each carry distinct emphasis and intrinsic limitations: laser-electron-beam collision concepts demand extremely spatiotemporal synchronization precision~\cite{li2020prl,tang2020plb,wan2020prres,wu2025cp}, rendering experimental implementation difficult; the transverse ponderomotive force of conventional Gaussian lasers drives electron divergence, producting $\gamma$ rays with large divergence angles~\cite{xue2020mre,qian2026njp}; vortex laser pulses can produce collimated attosecond $\gamma$-ray sources, but their output typically takes the form of attosecond pulse trains~\cite{zhu2018apl,sun2025prapp}, which cannot deliver the isolated pulses needed for undisturbed, single-shot observations~\cite{Chini2014np,Zhang2020prl}. Consequently, identifying a robust, all-optical mechanism that can simultaneously generate an isolated, collimated, high-brilliance, highly polarized, attosecond $\gamma$-ray beam remains a critical open challenge.

\begin{figure}[t]	 
	\centering
	\setlength{\abovecaptionskip}{-0.05cm}
	\setlength{\belowcaptionskip}{-0.5cm}	
	\includegraphics[width=1.0\linewidth]{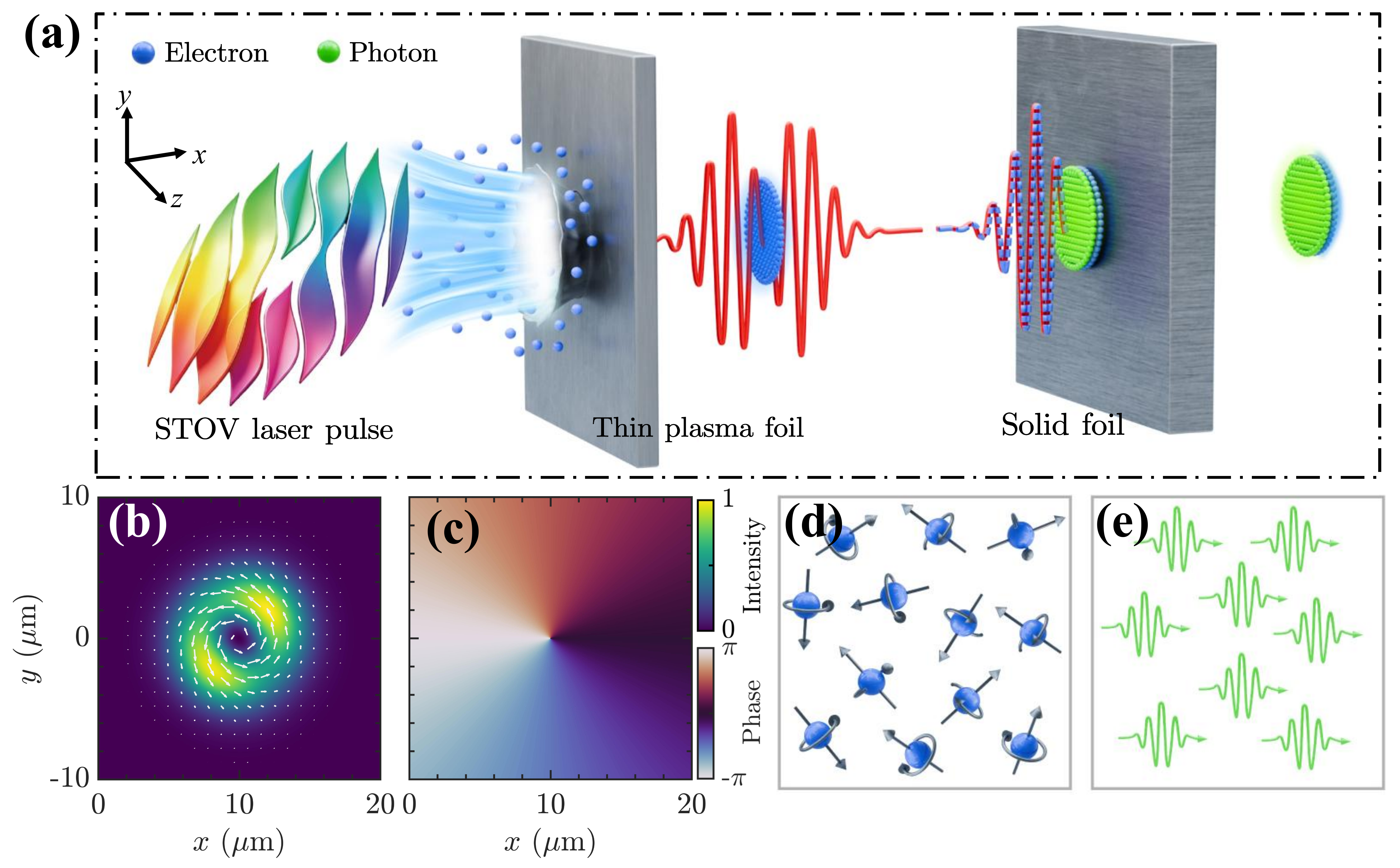}
	\caption{(a) Schematic of isolated $\gamma$-ray beam generation via LP STOV laser-plasma interaction. The solid red and dotted blue curves represent the transverse electric field profiles along the axes of the incident and reflected laser pulses, respectively. (b) Normalized intensity distribution of an LP STOV laser pulse in the $x$-$y$ plane, with the white arrows indicating the circulating momentum flux. (c) Theoretical vortex phase distribution of the STOV pulse in the $x$-$y$ plane. (d) Electron spin and (e) photon polarization distribution.}
	\label{Schematic}
\end{figure}

\begin{figure}[t]	
	\centering
	\setlength{\abovecaptionskip}{-0.1cm}
	\setlength{\belowcaptionskip}{-0.5cm}	
	\includegraphics[width=\columnwidth]{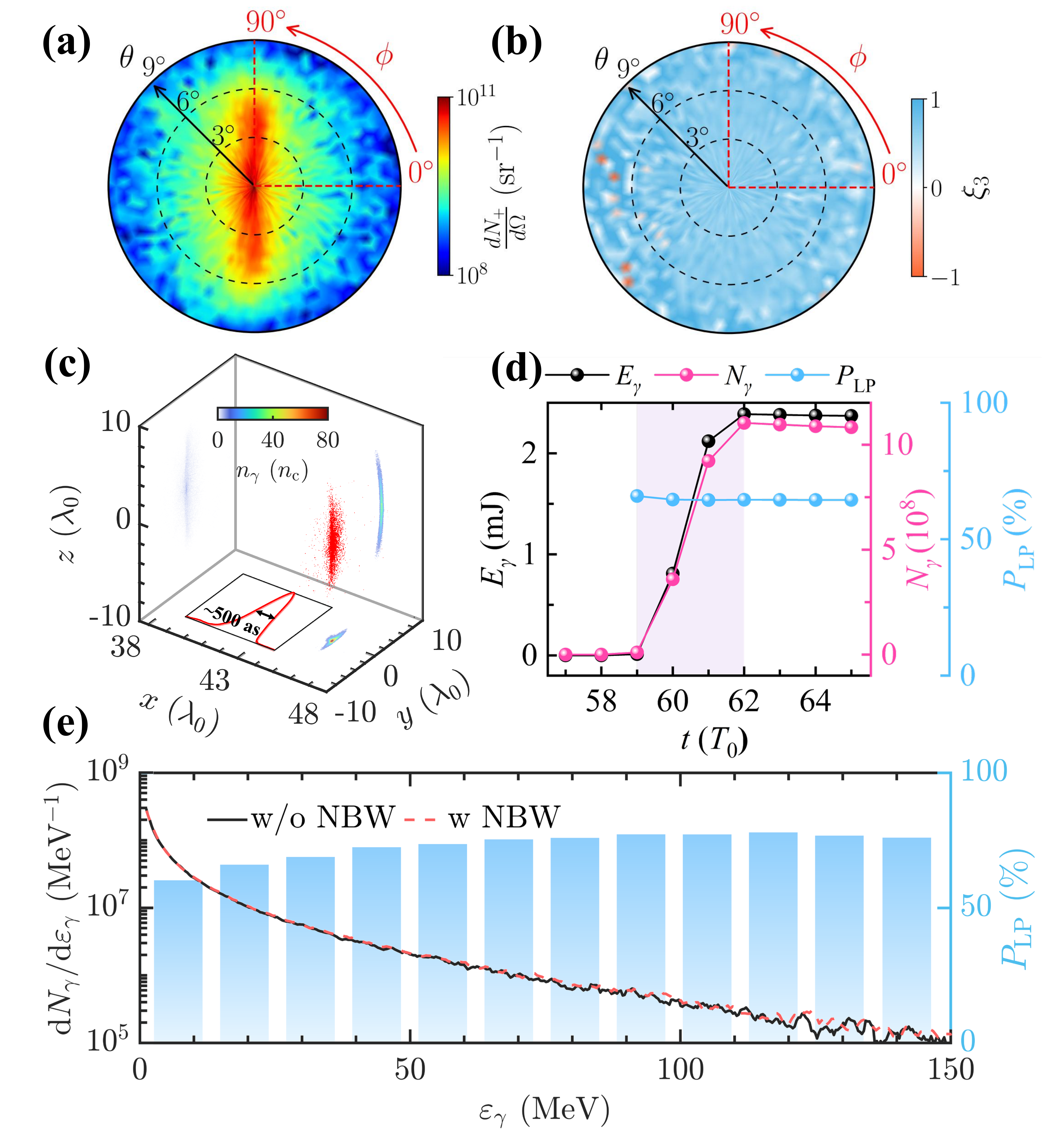}
	\caption{Angle-resolved distributions of (a) number density and (b) linear polarization degree for all emitted $\gamma$ photons in an isolated $\gamma$-ray beam at $t=62T_0$. (c) 3D iso-surface distribution of the photon number density, together with its distributions in the $x$-$y$, $x$-$z$, and $y$-$z$ planes. The inset in the $x$-$y$ plane shows the duration of the isolated $\gamma$-ray beam. (d) Temporal evolution of the photon energy $E_\gamma$, yield $N_{\gamma}$, and linear polarization degree $P_{\rm LP}$. The purple shaded region indicates the collision stage. (e) Energy spectrum of the isolated $\gamma$-ray beam with and without NBW, and the linear polarization degree distribution for different energy intervals at $t=62T_0$.}
	\label{Photon}
\end{figure}

In this Letter, we propose a scheme for isolated, collimated, high-brilliance, polarized attosecond $\gamma$-ray beams via the interaction of a linearly polarized (LP) STOV laser with plasma foils [Fig.~\ref{Schematic}(a)]. Carrying transverse orbital angular momentum, the STOV pulse features a characteristic hollow intensity profile and a spatiotemporal vortex phase~\cite{Bliokh2015PP,Jhajj2016PRX,Chong2020np,Liu2024PI}, which fundamentally alter plasma dynamics [Figs.~\ref{Schematic}(b) and \ref{Schematic}(c)]. As the pulse impinges on a thin plasma foil, electrons are expelled by the ponderomotive force and subsequently captured at the spatiotemporal singularity, forcing them into a phase-locked, longitudinally dominated acceleration that forms an isolated, unpolarized high-energy electron bunch [Fig.~\ref{Schematic}(d)]. This bunch then collides head-on with the laser pulse reflected from another solid foil, emitting an isolated polarized $\gamma$-ray beam via nonlinear Compton scattering (NCS) [Fig.~\ref{Schematic}(e)]. Three-dimensional (3D) spin-resolved quantum electrodynamics (QED) particle-in-cell (PIC) simulations demonstrate that a STOV pulse with a peak intensity of $\sim7\times10^{21}$ W/cm$^2$ yields an isolated, collimated ($\sim1.5^{\circ}$) $\gamma$-ray beam with an average linear polarization degree of $\sim64.4\%$ and a duration of about 500 as [Fig.~\ref{Photon}]. Our approach is feasible with currently available or upcoming laser facilities, as it avoids the need for exact spatiotemporal synchronization, and is robust against the laser and target parameters.

\textit{Simulation methods and setups.---}We perform 3D QED-PIC simulations to investigate the generation of polarized $\gamma$-ray beams sketched above using the {\scshape epoch} code~\cite{arber2015ppcf}. The QED strength is characterized by the quantum invariant parameter $\chi_{\rm e}=\left|e \right|\hbar/(m_{\text{e}}^{3}c^4) |F_{\mu \nu}p^{\nu}|$, where $e$ is the electron mass, $\hbar$ is the reduced Planck constant, $m_{\rm e}$ is electron mass, $c$ is the speed of light in vacuum, $F_{\mu \nu}$ is the field tensor, and $p^{\nu}$ is the electron four-momentum. The dominant QED process, NCS including spin and polarization effects, is implemented via the Monte Carlo algorithm in the local constant field approximation~\cite{Duclous2010PPCF,Ridgers2014JCP,Gonoskov2015PRE}. With the mean axes chosen as quantization axes, the spin vector $\mathbf{S}$ of non-radiating electrons and the Stokes vectors $\boldsymbol{\xi}$ of non-decaying photons are also need to be updated. The code has been benchmarked against the polarized QED-PIC code {\scshape yunic}~\cite{song2021yunic,song2021njp,song2022prl}, additional implementation details are provided in the Supplemental Material (SM)~\cite{SM}. The polarization stage of emitted photons is characterized by the Stokes parameters ($\xi_1$, $\xi_2$, $\xi_3$), defined with respect to the orthogonal basis ($\hat{\mathbf{e}}_1$, $\hat{\mathbf{e}}_2$, $\hat{\mathbf{e}}_v$), where $\hat{\mathbf{e}}_1$ is the unit vector along the electron transverse acceleration, $\hat{\mathbf{e}}_v$ is the unit vector along the electron velocity, and $\hat{\mathbf{e}}_2=\hat{\mathbf{e}}_v\times\hat{\mathbf{e}}_1$. Since the emission angle of a relativistic electron is confined within $\sim1/\gamma_{\rm e}\ll1$, where $\gamma_{\rm e}$ is the electron Lorentz factor, emitted photons propagate predominantly along $\hat{\mathbf{e}}_v$. To determine the mean polarization of the $\gamma$-ray beam, the Stokes parameters of each photon must first be normalized to the same observation frame, i.e., rotate the Stokes parameters of each $\gamma$-photon from its instantaneous frame to the same observation, and then calculate the average Stokes parameters~\cite{li2020prl}.

The size of simulation box is $x\times y\times z$=$17\lambda_0\times20\lambda_0\times20\lambda_0$, with the corresponding cells of $850\times800\times800$. An LP STOV laser pulse with a duration of 30 fs is normally incident from the left boundary, with its electric field polarized along the $\hat{y}$-direction. The transverse electric field of the STOV laser pulse is $E_y=E_0[(\zeta^2+y^2)/w^2]^{|l|/2}\exp[-(\zeta^2+y^2+z^2)/w^2]\exp[i(-l\phi_{\rm st}+kx-\omega t)]$, where $E_0=a_0m_{\rm e}c\omega/e$ is the peak amplitude of the electric field, $a_0=75$ is the normalized laser amplitude (corresponding to the laser intensity $\sim7\times10^{21}$ W/cm$^{2}$), $\omega=2\pi c/\lambda_0$ is the laser frequency, $\lambda_0=cT_0=1$ $\mu$m is the laser center wavelength, $T_0$ is the laser cycle, $\zeta=ct-x$ is the longitudinal coordinate local to the laser, $w=4\lambda_0$ is laser focus spot size, $l=1$ is the topological charge, $\phi_{\rm st}=\arctan(y/\zeta)$ is the spatiotemporal azimuthal angle, and $k$ is the wave number. A dual plasma foils configuration is employed in the simulations. The first foil, located 2$\lambda_0$ from the left boundary with a thickness of 0.5$\lambda_0$, is a thin plasma layer with an electron density of $30n_{\rm c}$, serving as the electron source, where $n_\text{c}=m_\text{e}\omega^2/(4\pi e^2)\approx1.1\times10^{21}$ cm$^{-3}$ is the plasma critical density. The other foil, acting as a reflector, is a solid carbon foil with a density of $200n_{\rm c}$, positioned at $x=45\lambda_0$ with a thickness of $1\lambda_0$. For the thin foil, each cell contains 50 macro-particles for both electrons and protons, whereas for the solid foil, 8 macro-particles per cell are used for electrons and 4 for fully ionized C$^{6+}$ ions. Absorbing boundary conditions are applied to both fields and particles.

\textit{Properties of the $\gamma$-ray beam.---}The angle-resolved number density and average linear polarization of the emitted $\gamma$ photons are shown in Figs.~\ref{Photon}(a) and \ref{Photon}(b), respectively. The total photon yield $N_\gamma$ with $\varepsilon_\gamma\ge1$ MeV reaches $1.1\times10^9$, which is comparable to the electron number $N_{\rm e}$ in the electron bunch. This is in agreement with the analytical estimation $N_{\gamma}\sim\alpha_fa_0N_{\rm e}\tau'/T_0$~\cite{Ritus1985}, where $\tau'$ denotes the effective electron-laser interaction time. The beam divergence angle peaks at $\theta=\arctan(p_\bot/|p_x|)\approx1.5^\circ$, with a full width at half maximum (FWHM) of 2.5$^\circ$. More importantly, the linear polarization parameter satisfies $\xi_{3}>0$ for the vast majority of emitted $\gamma$ photons [Fig.~\ref{Photon}(b)], integrating over the angular distribution yields an average linear polarization degree of $P_{\rm LP}=\sqrt{\bar{\xi}_{1}^{2}+\bar{\xi}_{3}^{2}}\approx64.4\%$. Furthermore, the emitted $\gamma$ photons exhibit pronounced spatial localization, forming a structurally well-defined and isolated $\gamma$-ray beam with a duration of $\sim$500 as (FWHM), which is primarily determined by the electron bunch duration~\cite{li2015prl} [Fig.~\ref{Photon}(c)].

\begin{figure}[t]	
	\setlength{\abovecaptionskip}{-0.01cm}
	\setlength{\belowcaptionskip}{-0.3cm}
	\centering
	\includegraphics[width=1.0\linewidth]{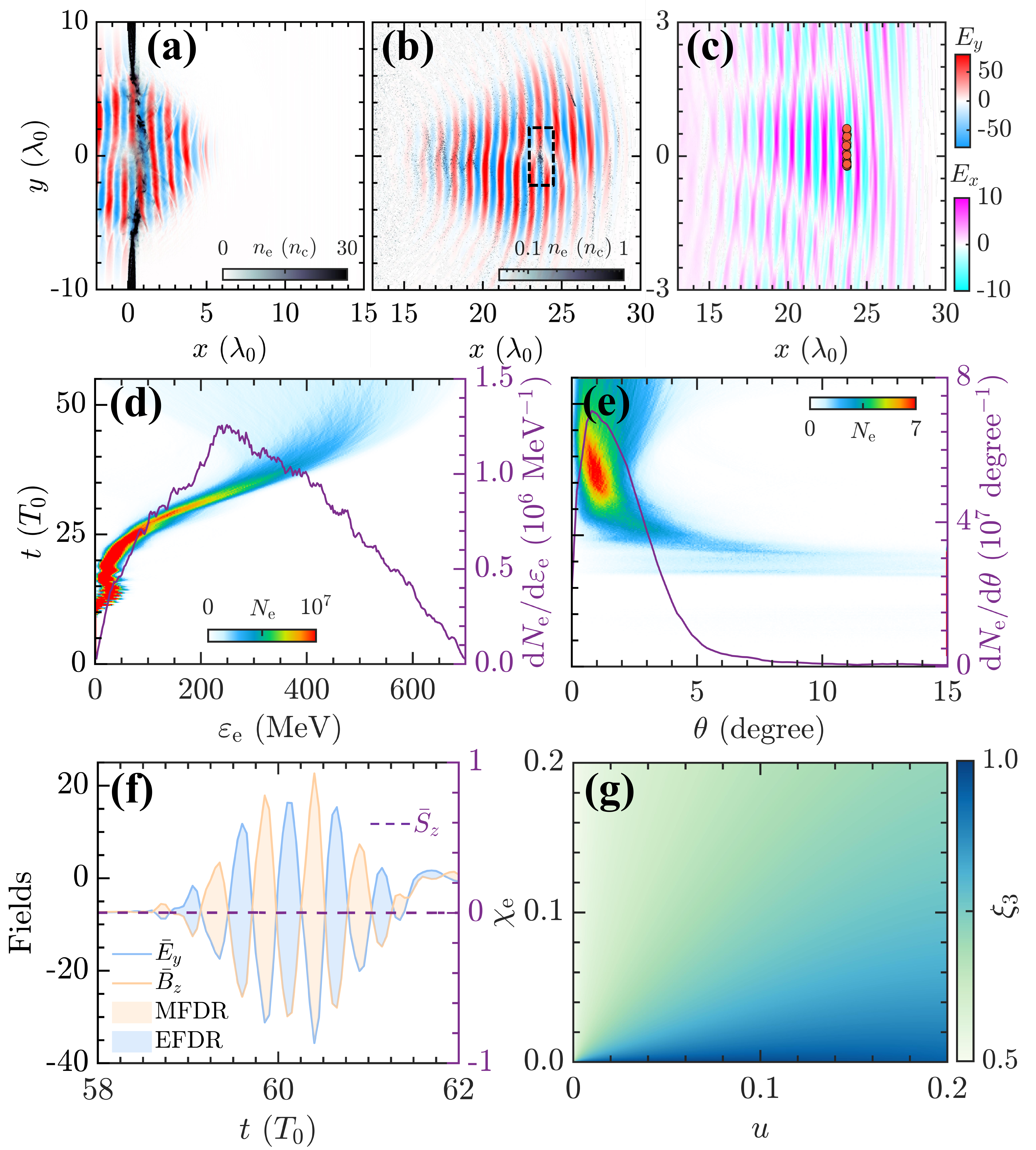}
	\caption{Distribution of $E_y$ and the electron number density in the $x$-$y$ plane at (a) $t=16T_0$ and (b) $40T_0$. (c) Enlarged view of (b), showing $E_x$ of the STOV pulse and typical electrons in the isolated electron bunch marked by the dotted box in (b). Both $E_y$ and $E_x$ are normalized by $m_\text{e}c\omega_0/e$. Temporal evolution of (d) the electron energy and (e) divergence-angle distribution in the isolated electron bunch. The energy spectrum and angle spectrum at $t=50T_0$ is shown by the solid purple curve. (f) Distributions of the average electric field $\bar{E}_y$, magnetic field $\bar{B}_z$, and transverse spin $\bar{S}_z$ experienced by the isolated electron bunch during the radiation stage. Yellow shaded regions indicate the magnetic-field-dominated regime (MFDR), and blue shaded regions indicate electric-field-dominated regime (EFDR). (g) Linear polarization Stokes parameter $\bar{\xi}_3$ as a function of $u$ and $\chi_{\rm e}$, obtained from Eq.~\ref{xi3_si0}.}		
	\label{Electron1}
\end{figure}

Driven by the NCS process, this isolated $\gamma$-ray beam is generated during interval from $58T_0$ to $62T_0$, during which the total energy and yield of $\gamma$ photons stabilize after an initial temporal increase, while the linear polarization remains nearly constant [Fig.~\ref{Photon}(d)]. Benefiting from both the high yield and the ultrashort duration, the total radiation flux of the isolated $\gamma$-ray beam with $\varepsilon_\gamma\ge1$ MeV reaches $\mathcal{F}_{\gamma}\approx2.2\times10^{24}$ s$^{-1}$, which is approximately two orders of magnitude higher than previously reported results~\cite{li2020prl,wu2025cp}. The photons exhibit an exponential energy spectrum with a cutoff energy approaching 150 MeV [Fig.~\ref{Photon}(e)], and the peak brilliance can reach $\sim10^{23}$ photons/s/mm$^2$/mrad$^2$/0.1\%BW at 1 MeV,  significantly surpassing earlier $\gamma$-ray sources~\cite{sarri2014prl,li2020prl,xue2020mre,Zhang2021hpl,Cui2025PRA}. In addition, the linear polarization degree increases from 60\% in the MeV regime to around 76\% near 150 MeV. This upward trend aligns well with the theoretical scaling where $P_{\rm LP}$ increases with $u=\varepsilon_\gamma/\varepsilon_{\rm e}$ within the investigated parameter regime [Fig.~\ref{Photon}(e) and Fig.~\ref{Electron1}(g)], where $\varepsilon_{\rm e}$ is the electron energy before the photon emission, and $\varepsilon_\gamma$ is the emitted photon energy. Notably, the effect of the nonlinear Breit-Wheeler (NBW) process on photon annihilation into electron-positron pairs is negligible [Fig.~\ref{Photon}(e)]. Such characteristics of broad energy coverage and high polarization fulfill the stringent requirements for exploring photonuclear interactions, such as quasideuteron excitation~\cite{LEE01101998} and giant dipole resonance~\cite{Chakrabarty2016}.

\textit{Electron dynamics and $\gamma$-ray polarizing mechanism.---}The mechanisms underlying the generation and polarization of the isolated $\gamma$-ray beam are illustrated in Fig.~\ref{Electron1}. The laser-foils interaction proceeds in three stages: electron injection, phase-locked acceleration, and $\gamma$-photon emission. Upon irradiating a thin plasma foil with an LP STOV laser pulse, a substantial number of electrons are expelled under the action of the laser ponderomotive force. The Lorentz force oscillatory component of the LP laser and the charge separation field induce coupled transverse and longitudinal electron oscillations. This initially confines the electrons near the foil surface, yielding a spatially divergent distribution [Figs.~\ref{Electron1}(a) and \ref{Electron1}(d)]. As the STOV pulse arrives, its spatiotemporal singularity progressively captures and guides these electrons toward the paraxial region [Fig.~\ref{Electron1}(b)]. The spatiotemporal singularity thus acts as a combined injector and phase filter, enabling simultaneous transverse convergence and longitudinal temporal selection for electrons that satisfying the phase-locked conditions.

Specifically, the transverse forces acting on the electrons are dominated by the laser electric field and the correspond Lorentz force. The transverse electric field simplifies to $E_y=E_0/w(\zeta-iy)\exp[-(\zeta^2+y^2+z^2)/w^2]\exp(i\psi^{'})$, where $\psi^{'}=kx-\omega t$. Paraxial confinement requires a net inward transverse force. At the spatiotemporal singularity, this convergence is governed by $F_y\approx-|e|E_y(1-v_x/c)\sim-(1-v_x/c)y\sin\psi^{'}$, establishing the transverse phase-locked condition $\sin\psi^{'}>0$. The longitudinal component, obtained from $\nabla\cdot\mathbf{E}=0$, is $E_x=E_0/(kw)\exp[-(\zeta^2+y^2+z^2)/w^2](-1+2y^2/w^2+2iy\zeta/w^2)\exp(i\psi^{'})$. In the paraxial region at the spatiotemporal singularity ($|y|\ll w$) [Figs.~\ref{Electron1}(b) and ~\ref{Electron1}(c)], the term $2y^2/w^2$ is negligible, reducing the field to $E_x\sim -\cos \psi^{'}$. Forward longitudinal acceleration demands $E_x<0$, which yields the longitudinal phase-locked condition as $\cos\psi^{'}>0$. Consequently, simultaneously transverse convergence and longitudinal acceleration thus restrict the phase to a narrow interval $\psi^{'}\in(0,\pi/2)$. This tight phase-locked window, spanning only one-quarter of a laser cycle, inherently limits the electron bunch duration to a few hundred attoseconds. 

During the acceleration stage, the injected electrons copropagate with the laser pulse [Fig.~\ref{Electron1}(c)], continuously gaining energy up to $\sim$700 MeV [Fig.~\ref{Electron1}(d)]. Concurrently, the divergence, initially broadened during injection, gradually  decreases and stabilizes, yielding a peak divergence angle of $0.8^{\circ}$, and a FWHM of $3^{\circ}$ at $t=50T_0$ [Fig.~\ref{Electron1}(e)]. Throughout this stage, $\mathbf{E}_{\bot}$ and $\boldsymbol{\beta}\times\mathbf{B}$ cancel exactly for the electrons, resulting in $\chi_{\rm e}\approx0$, which completely suppresses $\gamma$-photon emission. Our simulations demonstrate that an intense LP STOV laser pulse can generate an isolated ultra-relativistic electron bunch with a peak density of $\sim0.5n_{\rm c}$, a duration of 500 as, and a total charge of 110 pC. This mechanism provides a distinct route for generating isolated attosecond electron bunch, fundamentally different from Gaussian-laser-driven acceleration schemes [Fig. S4 in SM]~\cite{SM}.

The front of the LP STOV laser pulse is reflected by the planar solid foil, driving a head-on collision with the electron bunch. This triggers the NCS process, during which the quantum parameter $\chi_{\rm e}\approx2\gamma_{\rm e}|\mathbf{E}_{\bot}|/E_{\rm cr}$ rapidly exceeds 0.1, marking the onset of efficient isolated $\gamma$-ray beam generation, where $E_{\rm cr}=m_{\rm e}^2c^3/(e\hbar)\approx1.3\times10^{18}$ V/m is the QED critical field strength. Throughout the emission stage, the electron bunch remains confined in a symmetric laser field configuration, so that its net spin polarization vanishes [Fig.~\ref{Electron1}(f)]. According to spin-polarization-resolved NCS theory, the photon emission probability is given by ${\rm d}^2W_{\rm rad}/({\rm d}u{\rm d}t)=C_{\rm rad}(F_0+\xi_1F_1+\xi_2F_2+\xi_3F_3)/4$~\cite{li2020prl,chen2022PRD}, where $C_{\rm rad}=(\alpha_f m_{\rm e}^2c^4)/(\sqrt{3}\pi\hbar\varepsilon_{\rm e})$, and the functions $F_0$, $F_1$, $F_2$, $F_3$ are defined in SM~\cite{SM,Li2019PRL}. For such an unpolarized electron bunch ($\bar{\mathbf{S}}=0$), the average Stokes parameters reduce to $\bar{\xi}_1=\bar{\xi}_2=0$ and $\bar{\xi}_3>0$, indicating that the isolated $\gamma$-ray beam is predominantly LP along the $\hat{y}$-direction, with $\bar{\xi}_3$ explicitly expressed as
\begin{equation}
	\bar{\xi}_3=\frac{K_{2/3}(y)}{\left(u^2-2u+2\right) /(1-u)K_{2/3}(y)-\text{Int}K_{1/3}(y)}.
	\label{xi3_si0}
\end{equation}
Here $K_{\nu}(y)$ is the modified Bessel function of the second kind with a non-integer factor $\nu$, $\text{Int}K_{1/3}(y) =\int_y^{\infty}{K_{1/3}\left( x \right)}\text{d}x$, and $y=2u/[3(1-u)\chi_{\rm e}]$. Crucially, $\bar{\xi}_{3}$ is uniquely determined by $u$ and $\chi_{\rm e}$. Within the interval $0<u<0.2$ covered by our scheme, $\bar{\xi}_3$ decreases with increasing $\chi_{\rm e}$ at a fixed $u$, but increases with $u$ for a given $\chi_{\rm e}$ [Fig.~\ref{Electron1}(g)], elucidating the numerical trends observed in Fig.~\ref{Photon}(e).

\begin{figure}[t]	
	\centering	
	\setlength{\abovecaptionskip}{-0.05cm}
	\setlength{\belowcaptionskip}{-0.2cm}	
	\includegraphics[width=1.0\linewidth]{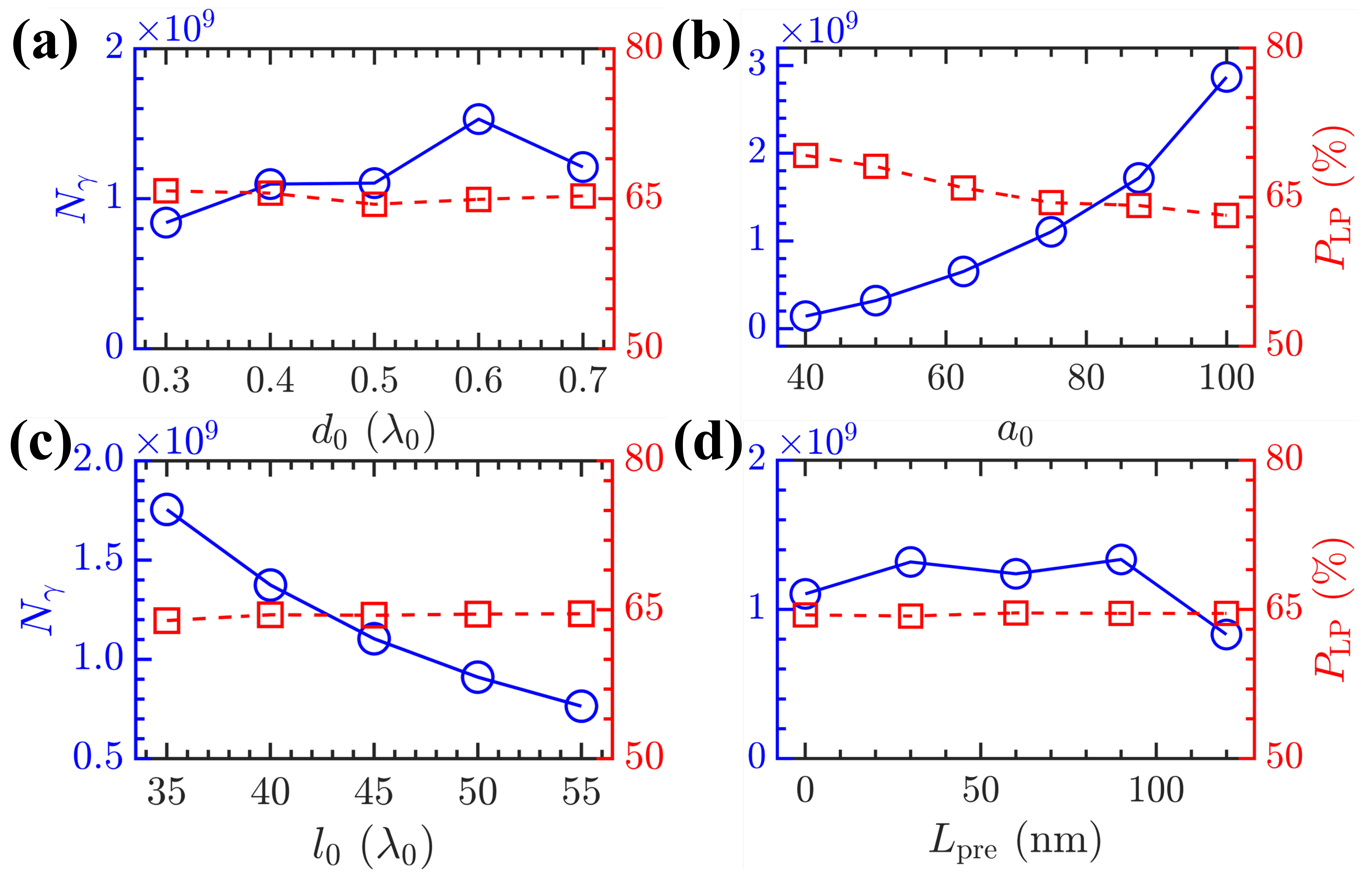}
	\caption{The yield $N_{\gamma}$ and averaged linear polarization degree $P_{\rm LP}$ of the isolated $\gamma$-ray beam vs (a) the thin plasma foil thickness $d_0$, (b) the laser peak intensity $a_0$, (c) the solid foil position $l_0$, and (d) the preplasma scale length $L_{\rm pre}$, respectively.}
	\label{robustness}
\end{figure}

\textit{Robustness and Parameter Dependence.---}To validate the experimental feasibility of our scheme, we investigate the influence of laser and target parameters on the photon number $N_\gamma$ and linear polarization degree $P_{\rm LP}$ of the isolated $\gamma$-ray beam, as summarized in Fig.~\ref{robustness}. In this scheme, the stable generation of an isolated $\gamma$-ray beam inherently depends on the formation and acceleration of an isolated electron bunch. The critical condition requires the laser radiation pressure to overcome the electrostatic space-charge field of the ions, satisfying $a_0>\pi(n_{\rm e}/n_{\rm c})(d/\lambda_0)$~\cite{Macchi2013RMP,Shen2021PRE}. To maintain a constant target areal density and systematically evaluate the robustness of this scheme, we adopt the scaling relation $a_0=5(n_{\rm e}/n_{\rm c})(d/\lambda_0)$. At a fixed laser intensity, as the thin plasma foil thickness $d_0$ increases, the photon yield $N_{\gamma}$ exhibits a non-monotonic trend, first increasing and then decreasing, while the $P_{\rm LP}$ remains virtually constant [Fig.~\ref{robustness}(a)]. The behavior occurs because an overly thin foil, corresponding to an excessively high density, generates a strong electrostatic field that suppresses electron injection. Conversely, a foil that is too thick implies a low density, triggering premature laser transmission that reduces the number of trapped electrons, this decreases $N_{\gamma}$ during the NCS process while leaving $\chi_{\rm e}$ nearly unaffected, thereby maintaining high linear polarization. 

Increasing the laser intensity $a_0$ significantly enhances $N_{\gamma}$ but degrades $P_{\rm LP}$ [Fig.~\ref{robustness}(b)]. This is attributed to the elevated $\chi_{\rm e}\propto a_0\gamma_{\rm e}$ under higher $a_0$, which strengthens the NCS process and enhances the photon emission probability at the expense of $P_{\rm LP}$. As the reflection foil position $l_0$ shifts backward, $N_{\gamma}$ gradually decreases, whereas $P_{\rm LP}$ improves slightly [Fig.~\ref{robustness}(c)]. This is driven by spatiotemporal diffraction effects during the propagation of the STOV laser pulse, which causes intensity attenuation and envelope deformation. These effects not only weaken the phase-locked acceleration and reduce the charge of the trapped electron bunch, but also lower the peak intensity and $\chi_{\rm e}$ during the subsequent collision, ultimately resulting in a reduced $N_{\gamma}$ and a minor increase in $P_{\rm LP}$. Furthermore, preplasmas generated by laser prepulse are unavoidable and also adjustable in real experiments. For scale lengths $L_{\rm pre}<90$ nm, the $N_{\gamma}$ increases slightly with $L_{\rm pre}$, indicating that a moderated scale length optimizes laser-plasma energy coupling and electron injection. However, an excessively long scale length leads to premature dissipation of laser energy, reducing the number of effectively accelerated electrons and consequently decreasing $N_{\gamma}$. Throughout this process, the $P_{\rm LP}$ remains insensitive to the $L_{\rm pre}$ because $\chi_{\rm e}$ in the collision region is largely unaltered [Fig.~\ref{robustness}(d)]. Additionally, we investigated the configuration involving a tilted solid reflection foil, demonstrating that high-quality isolated and polarized $\gamma$-ray beam can still be generated with negligible impact [Fig. S5 in SM]~\cite{SM}.

\textit{Conclusions.---}In conclusion, we investigate a novel all-optical method to generate an isolated, collimated, high-brilliance, polarized attosecond $\gamma$-ray beam via a single-shot LP STOV laser-foils interaction. The essence of this scheme lies in the unique spatiotemporal topology of the STOV pulse. This spatiotemporal singularity traps and accelerates electrons that subsequently collide with the reflected pulse to emit photons, thereby intrinsically unifying electron injection, acceleration, and photon emission within a single structured light field, enabling simultaneous control over the linear polarization degree, collimation, and attosecond temporal structure of the $\gamma$-ray beams. Such high-quality $\gamma$-ray source opens new prospects for attosecond positron beams generation under extreme conditions, and for polarization-sensitive nuclear physics investigations.

\setParDis
\textit{Acknowledgments.---}We thank Dr. H. Zhang for fruitful discussions. This work was supported by the National Natural Science Foundation of China (Grant No. 12575262, 12388102, and 12405285), the National Key R\&D Program of China (Grant No. 2025YFF0515101), the Shanghai Municipal Science and Technology Major Project, the Strategic Priority Research Program of the Chinese Academy of Sciences (Grant No. XDA0380000), the Natural Science Foundation of Shanghai (Grant No. 25JD1404100), and the China Postdoctoral Science Foundation (Grant No. 2025M783448). 
\setParDef

\setParDis
\textit{Data availability.---}The data supporting the findings of this study are presented in the figures within the article. The raw-underlying data that support the results are available from the corresponding authors upon reasonable request.
\setParDef

\bibliography{sample}

\end{document}